# A Derivation of the Power Dissipation Index using Peak Virtual Power.

CE Neal-Sturgess.

Emeritus Professor of Mechanical Engineering

The University of Birmingham.

**ABSTRACT**:

It is shown that the physical damage resulting from severe weather phenomenon such as hurricanes and tornadoes encapsulated in various scales such as Saffir-Simpson, TORRO and the Enhanced Fujita Scale, and measured by the Power Dissipation Index can be predicted using the concept of the Peak Virtual Power. This corresponds to the maximum rate of Entropy production, which is considered to be the basic physical principle underlying the damage.

**INTRODUCTION**:

Physical damage to buildings, infrastructure, trees or humans is thought to be an ever-increasing problem due to the extremes of weather forecast through the effects of Global warming [IPCC 2014]. Although the main causes of death in such events is flooding, physical damage is also important. Severe weather has been measured using various empirical scales since the introduction of the Beaufort Wind Force Scale in 1805 [NML Fact Sheet 6]. There are basically two types of severe weather scale, namely wind speed scales, of which all but Beaufort Scale 12 are, and damage scales such as the Saffir-Simpson hurricane scale [NHC 2009], TORRO [Meaden 2004] and the Enhanced Fujita Scale for Tornadoes [Murphy 2018]. Emanuel [2005] introduced the concept of the Power Dissipation Index (PDI) based on the cube of the windspeed, which is a well-known measure of wind power [Bird], integrated over time. This is a good measure with some correlation but what is missing is the physics underlying the measure; that is the purpose of this paper.

Damage caused by Hurricanes is damage to physical objects or structures may be represented as elastic-plastic or visco-plastic materials, and the damage may be viewed as dissipative mechanical processes modelled using Continuum Damage Mechanics (CDM).

Modelling involves predicting the future, and all that is known of the future is that Entropy must increase, as a consequence of the Second Law of Thermodynamics. Irreversible thermodynamics features the Clausius-Duhem Inequality, which is a consequence of rate dependent form of the Second Law of Thermodynamics [Prigogine 1997; Jou, Casa-Vazquez et al. 1993). This equation gives a formalism for deriving constitutive equations for irreversible dissipative processes, which since the 1980's has been applied to a number of flow and damage problems in CDM (Krajcinovic 1983; Krajcinovic and Lamaitre 1986; Chaboche 1988; Chaboche 1988; Lamaitre and Chaboche 1990). The concept of Peak Virtual Power has been used extensively to accurately formulate various theorems in the mechanics of materials and has recently been applied to human injuries in blunt impact trauma (Sturgess 2001). Much of the theory has been published previously [Sturgess 2001] but is included here for completeness. It should be realised that PVP is a principle, not a formula, and the form it takes is dependent on the application.

**THEORY**:

If the process is chemically and electrically neutral, the Clausius-Duhem Inequality for small deformation of incompressible bodies is (Jou, Casa-Vazquez et al. 1993):

$$\sigma : \dot{\varepsilon} - \rho(\dot{f} + s\dot{T}) - \frac{1}{T} q . \nabla T \geq 0 \tag{1}$$

where: $\sigma$ = Stress Tensor, $\dot{\varepsilon}$ = Total Strain Rate Tensor, $\rho$ $\rho$= Mass Density, $f$ = Helmholtz Free Energy, s = Entropy, T = Absolute Temperature, q = Heat Flux Vector

This is a Tensor equation as it must be invariant to transformations. Proceeding according to (Krajcinovic and Lamaitre 1986) assuming that the total strain tensor can be decomposed into the elastic and plastic (reversible and irreversible) strain tensors as:

$$\varepsilon = \varepsilon^e + \varepsilon^p \tag{2}$$

This decomposition applies to both elastic-plastic and visco-plastic materials (Lamaitre and Chaboche 1990). The appropriate form of the Helmholtz Free Energy (the constitutive relationships) (Jou, Casa-Vazquez et al. 1993) may be described as a function of the following variables:

$$f = f(\varepsilon^e, D_{ij}, T) \tag{3}$$

where $\varepsilon^e$ = the elastic strain tensor
$D_{ij}$ = the "Damage Tensor"

Now differentiating (3) and substituting in (1) gives the specific rate of entropy production ($\dot{\Phi}$) as:

$$\rho\dot{\Phi} = \sigma:\dot{\varepsilon}_p - \rho\left(\frac{\partial f}{\partial D_{ij}}\right)\dot{D}_{ij} - \frac{1}{T}q.\nabla T \geq 0 \tag{4}$$

This equation can be developed to derive the Maxwell, Huber, von-Mises, Hencky yield criteria, and the fundamental equations of Fracture Mechanics (Murakami 1988; Murakami and Kamiya 1997; Li 1999)

For a mechanical system, assuming the contribution of heat flux can be decoupled (Lamaitre and Chaboche 1990),

$$\sigma:\dot{\varepsilon}^p \geq \rho\left(\frac{\partial f}{\partial D_{ij}}\right)\dot{D}_{ij} \quad \text{or} \quad \sigma:\dot{\varepsilon}^p \propto \dot{D}_{ij} \tag{5}$$

If sufficient data is available the form of the self-consistent constitutive equations and the damage evolution function can be determined, see (Krajcinovic 1983; Krajcinovic and Lamaitre 1986; Chaboche 1988; Chaboche 1988; Murakami 1988; Lamaitre and Chaboche 1990) equation (5) could be integrated. However, for general systems these functions are not known.

Equation (5) shows that the Virtual Power, i.e. product of stress and strain-rate, is proportional to the rate of damage production provided this is dominated by mechanical dissipation. Equivalent stress and strain must be used as they are Invariant.

From which if equations (5) and (6) are compared, it can be that virtual power is proportional to the rate of damage production which is also proportional to the rate of specific entropy production; this was also identified by Martin (Martin 1975). However, power is not necessarily unique, but "Peak Virtual Power" (PVP) is unique and will be defined as:

$$PVP = \frac{1}{\rho}\left[\frac{\partial(U/V)}{\partial t}\right]^{\Uparrow} = \left|\sigma_{ij}\dot{\varepsilon}_{ij}^{p}\right|^{\Uparrow} \quad (6)$$

For a given process, over a given time interval, the amount of damage (severity) will be proportional to the rate of damage production times the time interval, a "rate", "dosage", or "exposure" criterion then becomes: Type equation here.

$$Severity\ of\ Damage \propto D \propto \int_0^t \dot{D}_{ij} dt \propto PVP \quad (7)$$

In a hurricane or tornado, the forcing function is the wind power, which is proportional to the wind velocity cubed [Bird], as shown below:

The mass of air passing through area A in small time dt is $\rho A dx$,

    Where:      ρ is mass/volume

                   dx is distance travelled in time dt,

                   so $dx = vdt$

$$\text{and kinetic energy} = \frac{1}{2}mv^2 = \frac{1}{2}\rho A v^3 dt \quad (8)$$

The rate of energy dissipation is:

$$Power = \frac{1}{2} \rho A v^3 \qquad (9)$$

Therefore, in this system the Peak Virtual Power is

$$PVP \propto \hat{v}^3 \qquad (10)$$

The Beaufort Wind Velocity Scale as its name implies is a wind speed scale, devised essentially for mariners in the 19th century to better describe the forces exerted by the wind on sailing ships. However, Beaufort Scale 12 is classified as a hurricane, and is the only point on the Beaufort Scale which is a measure of damage. For the purposes of this paper this is the only Beaufort point that will be considered.

The Saffir-Simpson Hurricane Scale is a scale classifying most tropical cyclones occurring in the western hemisphere that exceed the intensities of tropical depressions and tropical storms, and so become Hurricanes. The categories into which the scale divides Hurricanes are distinguished by the intensities of their respective sustained winds. The classifications are intended primarily for use in measuring the potential damage and flooding a hurricane will cause upon landfall. For the purposes of this paper the damage to physical structures is the focus, the flooding is consequential.

The Saffir-Simpson Scale is [Wikipedia, Nov. 2020]:

**Saffir-Simpson Hurricane Scale**

| Category | Wind speed mph (km/h) | Storm surge ft (m) |
|---|---|---|
| 5 | ≥156 (≥250) | >18 (>5.5) |
| 4 | 131–155 (210–249) | 13–18 (4.0–5.5) |
| 3 | 111–130 (178–209) | 9–12 (2.7–3.7) |
| 2 | 96–110 (154–177) | 6–8 (1.8–2.4) |
| 1 | 74–95 (119–153) | 4–5 (1.2–1.5) |

The Enhanced Fujita Scale, or EF Scale, is the scale for rating the strength of tornadoes in the US, and is based on the damage they cause, i.e. it is a *damage* scale and is [Edwards, 2013].

| Category | Wind speed | Damage |
|---|---|---|
| Unknown | | No surveyable damage |
| EF0 | 65–85 mph | Light damage |
| EF1 | 86–110 mph | Moderate damage |
| EF2 | 111–135 mph | Considerable damage |
| EF3 | 136–165 mph | Severe damage |
| EF4 | 166–200 mph | Devastating damage |
| EF5 | >200 mph | Incredible damage |

Implemented in place of the Fujita Scale, which was introduced in 1971, it began operational use in 2007 [Edwards, 2013].  The scale has the same basis as the original Fujita scale, with six categories from zero to five representing increasing levels of damage. It was revised to reflect better examinations of tornado damage surveys, so as to align wind speeds more closely with associated damage. Better standardizing and elucidating what was previously subjective and ambiguous. It also adds more types of structures as well as vegetation, expands degrees of damage, and better accounts for variables such as differences in construction quality.

The TORRO tornado intensity scale (or T-Scale) is a scale measuring tornado intensity between T0 and T11. It was developed by Terence Meaden of the Tornado and Storm Research Organisation (TORRO) in the UK as an extension of the Beaufort Scale [ Meaden, 2004].  TORRO claims it differs from the Fujita Scale in that it is claimed to be purely wind speed scale, whereas the Fujita scale relies on damage for classification, but in practice, damage is utilised almost exclusively in both systems to infer intensity. That is because such a proxy for intensity is usually all that is available; although users of both scales would prefer direct, objective, quantitative measurements. The scale is primarily used in the United Kingdom whereas the Fujita scale is the primary scale used in North America, Europe, and the rest of the world.

**Results**:

Taking the mean wind speed for each category on the various scales and plotting the mean wind speed cubed (PVP) against the various categories gives Fig.1 below.  In this figure to express all the scales on one set of axes for the Enhanced Fujita Scale what is plotted is the EFS + 1.  In the case of the TORRO scale only the first six categories are plotted, although the scale goes up to category 10, to compare with the Saffir-Simpson and the Enhanced Fujita scales.

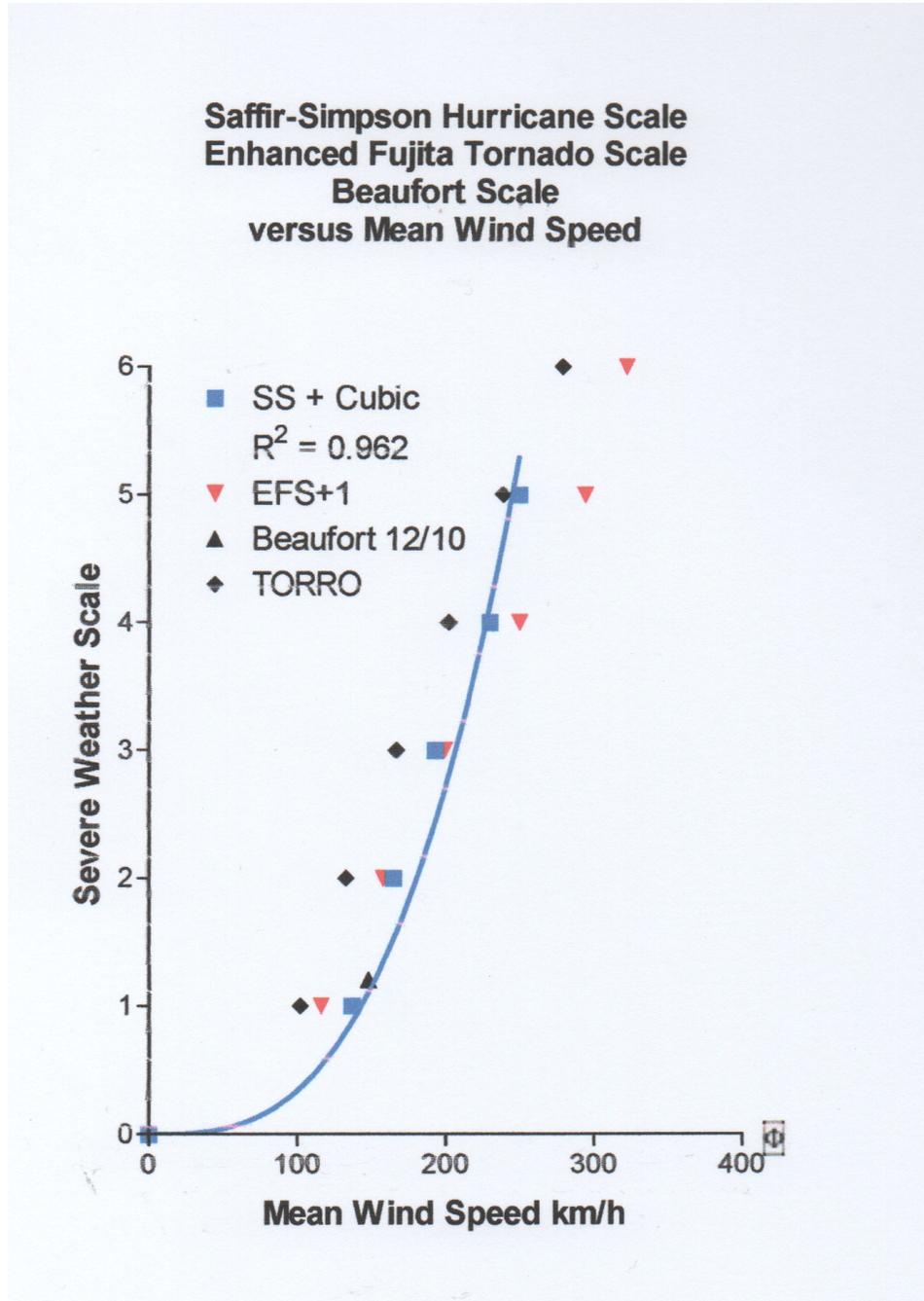

It can be seen from Fig.1 that the Saffir-Simpson scale shows good correlation with Peak Virtual Power. This is shared with the first five categories of the Enhanced Fujita Scale. It can also be seen that the TORRO scale also follows the same trend, although there is a shift downward in the wind speed associated with a given category. Finally, it is shown that Beaufort Scale point 12 (hurricane), plotted as Beaufort scale divided by 10 to fit on the same axes, also fits the same curve.

## Conclusion:

It is shown that Power Dissipation Index can be derived using Peak Virtual Power and the phenomenological Saffir-Simpson Scale for the damage potential of a Hurricane, the Enhanced Fujita Scale and the TORRO scale for the damage rating of Tornadoes, together with Beaufort Scale 12 are also modelled accurately by the concept of Peak Virtual Power up to a mean wind speed of 250 km/h. As PVP is proportional to the maximum rate of Entropy production, it is considered that this is the underlying physical basis for the severe weather categorisations in current use.